\documentclass[%
preprint,
showpacs,preprintnumbers,
 nobibnotes,
 amsmath,amssymb,
 aps,
 prl,
]{revtex4-1}

\usepackage{%
 graphicx,
 hyperref,
 dcolumn,
}

\begin{document}

\title{Atomic Bloch-Zener Oscillations and St\"uckelberg Interferometry
in Optical Lattices}

\author{Sebastian Kling}
\email{kling@iap.uni-bonn.de}
\author{Tobias Salger}
\author{Christopher Grossert}
\author{Martin Weitz}

\affiliation{Institut f\"ur Angewandte Physik der Universit\"at Bonn,
Wegelerstr. 8, 53115 Bonn}

\date{\today}

\begin{abstract}

We report on experiments investigating quantum transport and band 
interferometry of an atomic Bose-Einstein condensate in an optical
lattice with a two-band miniband structure, realized with a
Fourier-synthesized optical lattice potential. Bloch-Zener
oscillations, the coherent superposition of Bloch oscillations and
Landau-Zener tunneling between the two bands are observed. When
the relative phase between paths in different bands is varied, an
interference signal is observed, demonstrating the coherence of
the dynamics in the miniband system. Measured fringe patterns of
this St\"uckelberg interferometer allow to interferometrically map
out the band structure of the optical lattice over the full
Brillouin zone.

\end{abstract}


\maketitle

Transport properties of a quantum object in a periodic potential
are crucially determined by the particle's band structure
\cite{AshcroftMermin}. For example, Bloch oscillations, in which a
particle subjected to a uniform force in a periodic potential
performs an oscillatory rather than a uniformly accelerated
motion, can be well described by the dynamics within a single band
of the Bloch spectrum \cite{Feldmann92, BenDahan96}. For systems
with two bands energetically separated from energetic higher bands
(miniband structure), as can be realized by imposing a
superlattice structure onto a usual sinusoidal lattice potential,
Bloch-Zener oscillations, a characteristic sequence of
Bloch-oscillations and Landau-Zener transitions, have been
predicted to occur when a constant force is applied
\cite{Breid06a, Breid06b}. The avoided crossing between the two
minibands can furthermore act as a coherent beam splitter when
partial Landau-Zener tunneling between the subbands occurs. When
an avoided crossing is used first to coherently split up and
subsequently to recombine atomic wavepackets, we expect to be able
to observe an interference pattern, which is analogous to the
St\"uckelberg oscillations long known in collisional atomic
physics \cite{Stueckelberg32, Nikitin84}. Experimentally,
Bloch-Zener oscillations have been observed for light waves in
waveguide arrays, while the phase coherence between interfering
path has not been explicitly verified \cite{Dreisow09}.
St\"uckelberg interference based on two partial Landau-Zener
transitions has been observed with Rydberg atoms \cite{Yoakum92},
superconducting systems \cite{Oliver05, Sillanpaa06}, and in
ultracold molecular physics where this method was applied to
measure Feshbach molecular levels \cite{Mark07}. Recent
theoretical work has discussed the possibility to simulate
quasirelativistic effects in a two-band cold atom system, which in
the presence of interactions can be described by a nonlinear Dirac
equation \cite{Merkl10}.

Here we report on the observation of Bloch-Zener oscillations with
ultracold atoms in an optical lattice with a two-band miniband
structure. We have also observed an interference signal based on
two partial Landau-Zener transitions between Bloch bands of the
optical lattice. From measured fringe patterns of this
St\"uckelberg interferometer, the energetic splitting between the
bands at an arbitrary value of the atomic quasimomentum can be
determined. This realizes a novel method to interferometrically
map out the band structure of an optical lattice.

Before proceeding, we note that band structure determinations have
a long history in the solid state physics community, and powerful
methods here include neutron scattering and both optical and X-ray
spectroscopy \cite{AshcroftMermin}. For cold atoms in optical
lattices, vibrational frequencies and the splitting between bands
at the position of the gaps can be readily determined by the well
established techniques of parametric heating \cite{Friebel98},
Rabi-oscillations at the gaps \cite{Mellish03} and Landau-Zener
tunneling \cite{Oberthaler06}. The full band structure of an
optical lattice could in principle be determined by Bragg
spectroscopy, but this requires a continuous change of the
angle between the driving laser beams \cite{Clement09}.

\begin{figure}
\includegraphics{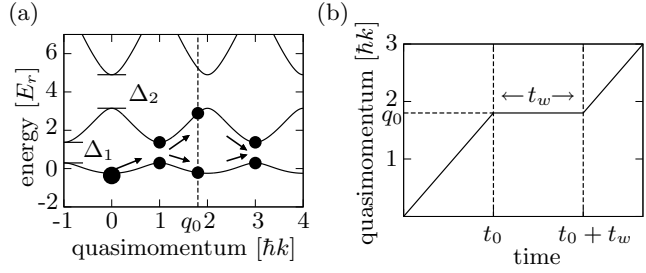}
\caption{%
(a) Band structure of the Fourier-synthesized optical lattice
potential along with a scheme of the St\"uckelberg interferometer.
In the lattice frame, atoms are accelerated over the first
bandgap, where partial Landau-Zener tunneling creates a coherent
superposition of wavepackets in the lowest two Bloch-bands. To
induce a controllable phase shift between paths, the acceleration
can be stopped at a quasimomentum $q_0$ (with $\hbar k\le q_0 \le
3\hbar k$) for a waiting time $t_w$. The acceleration then
continues until this bandgap is again reached at a quasimomentum
of $3\hbar k$ to close the atom interferometer. The parameters used 
for the band structure were $V_1 = 2.7$ $E_r$, $V_2 = 3.2$ $E_r$, 
and $\phi = 0^\circ$. (b) Variation of the quasimomentum with time 
for realizing the St\"uckelberg interferometer. 
} \label{fig1}
\end{figure}

Our experiment uses lattice potentials realized by superimposing a
conventional standing wave lattice potential of $\lambda/2$
spatial periodicity with a $\lambda/4$ periodicity lattice
realized with the dispersion of a multiphoton Raman process
\cite{Ritt06, Salger07}. The splitting between the lowest
energetic Bloch band and the first excited band is determined by
first order Bragg scattering of the standing wave lattice
potential, while the splitting between the first and the second
excited band is due to the interference of contributions of second
order Bragg scattering of the standing wave lattice potential and
that of first order Bragg scattering of the fourth order
($\lambda/4$-spatial periodicity) lattice. By choosing a
relatively large value of the amplitude of the multiphoton lattice
potential, the splitting between the first and the second excited
band can be made large so that the tunneling rate to higher bands
is small. In this way, a miniband structure with two closely
spaced subbands is prepared, in between the Landau-Zener
tunneling rate is large. In the following, $V_1$ and $V_2$ denote
the potential depths of the two lattice harmonics with spatial
periodicities $\lambda/2$ and $\lambda/4$ respectively, $\phi$ the
relative phase between lattice harmonics, and $k=2\pi/\lambda$ is
the optical wavevector. For a Fourier-synthesized lattice
potential of the form $V(z) = (V_1/2)\cos(2kz) +
(V_2/2)\cos(4kz+\phi)$, if $\Delta_1$ and $\Delta_2$ denote the
splittings between the ground and the first excited band, and the
first and the second excited band respectively (see
Fig.~\ref{fig1}a), $\Delta_2$ is maximised for a relative phase
between lattice harmonics of $\phi = 0$ \cite{Salger07}.
 If $\Delta_2 > \Delta_1$, as is desirable when
considering atom dynamics in the lowest two bands, the
corresponding bands are commonly called minibands. In general,
minibands emerge above a certain value of the ratio of the
potential depth of the harmonics: $V_2/V_1 > r(\phi, V_1, V_2)$.
Fig.~\ref{fig1}a shows the calculated band structure of such a
lattice for the experimental parameters used.

Our experiment starts with an atomic rubidium Bose-Einstein
condensate loaded into a lattice at rest relatively to the rest
frame of the atoms, so that atoms are transferred into the lowest
energy band at a quasimomentum $q(0) = 0$. Subsequently, the
lattice is accelerated relatively to the free falling atoms, which
is equivalent to the application of an external force $F$ to the
atoms, and the quasimomentum evolves in time to larger
quasimomenta, following $q(t) = q(0) + F\cdot t$. By the time that
the atomic wavepacket reaches the first bandgap, part of the
wavepacket experiences Landau-Zener tunneling through the gap
into the first excited Bloch band, while the remaining part
remains in the lowest band and is Bragg reflected. The
corresponding beam splitting process is visualized in the extended
band structure scheme shown in Fig.~\ref{fig1}a. The splitting
ratio can be controlled by tuning the size of the gap, which to
lowest order is determined by the magnitude of $V_1$. In this way,
a coherent superposition of wavepackets in the two different
subbands of the miniband structure is created. To shift the
relative phase between the two paths, the acceleration can be
stopped at some value of the quasimomentum $q_0$ (with $q_0 = q(t_0)$ and
$\hbar k\le q_0 \le 3\hbar k$) for a waiting time $t_w$. We expect that
the wavefunctions during this waiting time in the ground and first
excited subbands then evolve as:
\begin{eqnarray}
\label{tev}
\psi_1(q_0,t) &=& \psi_1(q_0,t_0)\exp{(-E_1(q_0)(t-t_0)/\hbar)}\,,\\ \nonumber
\psi_2(q_0,t) &=& \psi_2(q_0,t_0)\exp{(-E_2(q_0)(t-t_0)/\hbar)}\,,
\end{eqnarray}
where $E_1(q_0)$ and $E_2(q_0)$ denote the corresponding
eigenenergies at a quasimomentum $q_0$, so that a relative phase
between the subbands $\Delta\varphi (q_0,t_0+t_w)=
[E_2(q_0)-E_1(q_0)]t_w/ \hbar$ is accumulated in a time $t_w$.
Subsequently, the acceleration is continued, and at a
quasimomentum of $3\hbar k$ in the extended band scheme of
Fig.~\ref{fig1}a we again reach the bandgap between the ground and
the first excited Bloch band, where Landau-Zener tunneling acts
as a second beam splitter to recombine the two wavepackets and
close the atom interferometer.

The acceleration stops here at a quasimomentum of $3\hbar k$ and the lattice is
switched off, after which a time of flight image is recorded. 
Correspondingly, the lattice eigenstates at the position of the 
crossing are mapped onto the free atomic eigenstates. 
Depending on the relative phase
between wavepackets, atoms at the interferometer output will 
either be transferred into the first or the second diffraction order in
the far field image. As a function of the waiting time $t_w$ in
the acceleration sequence, we expect a sinusoidal fringe pattern
oscillating between the two different lattice diffraction orders.
Notably, as this fringe pattern oscillates at a frequency $\omega
= [E_2(q_0)-E_1(q_0)]/ \hbar$, the oscillation frequency allows to
interferometically determine the energetic splitting between
ground and first excited band of the Bloch spectrum at a given value
of the quasimomentum $q_0$. Fig.~\ref{fig1}b shows a scheme of the
variation of the lattice quasimomentum with time. It is clear that
from a variation of the quasimomentum at which the acceleration is
stopped, the complete spectrum $\omega(q)$ of the miniband
structure can be mapped out. We expect that this description is
valid when the Landau-Zener tunneling rate into higher bands is
small, as can be achieved for gap sizes $\Delta_2
> \Delta_1$ in the miniband structure. Otherwise, we expect a
reduced number of atoms at the interferometer output due to loss
into other diffraction orders.

\begin{figure}
\includegraphics{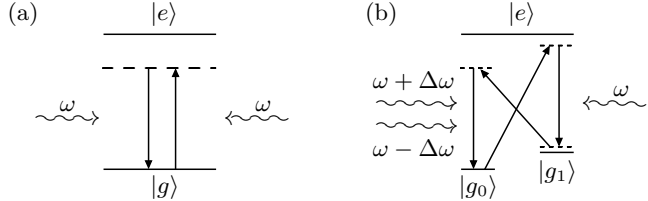}
\caption{%
(a) Virtual two-photon process in a standing wave lattice with
$\lambda/2$ spatial periodicity. (b) Virtual four-photon processes
contributing to a lattice potential with $\lambda/4$ spatial
periodicity, as the higher harmonic of the used
Fourier-synthesized lattice potential.}
\label{fig2}
\end{figure}

Bloch-Zener oscillations now refer to measurements where the relative phase
between the path between wavepackets was left at a constant value, as can be
reached by simply setting $t_w = 0$, (i.e. omitting the waiting time), and
monitoring the populations in the different bands versus time. In this case,
we expect that the mean atomic momentum performs a characteristic double
periodic motion with the two Bloch periods $T^{(1)}_B = 2\hbar k/F$ and
$T^{(2)}_B = 4\hbar k/F$ \cite{Breid06a}. This corresponds to a coherent
superposition of Bloch-oscillations and Landau-Zener tunneling in the
two-miniband structure.

Our experimental setup used to investigate ultracold rubidium
atoms in a Fourier-synthesized optical lattice is similar as
described previously \cite{Cennini03}. Briefly, an atomic rubidium
($^{87}$Rb) Bose-Einstein condensate is produced all-optically by
evaporative cooling of atoms in a CO$_2$-laser dipole trap. In the
final stages of evaporation, a magnetic field gradient is applied,
and this results in a spin-polarized condensate with $3\cdot10^4$
atoms in the $m_{\rm F} = -1$ component of the F = 1 hyperfine
ground state. The method used to produce a Fourier-synthesized
optical lattice potential for the atoms \cite{Ritt06}, with which
the described miniband structure can be achieved, is as follows.
For the fundamental spatial frequency of periodicity $\lambda/2$,
a usual standing wave lattice is used produced by two
counterpropagating beams of frequency $\omega$ (red detuned from
an atomic resonance). The well known processes contributing in a
quantum picture to this lattice potential arising from the
spatially varying ac-Stark shift are indicated in
Fig.~\ref{fig2}a. A lattice potential of spatial periodicity
$\lambda/4$, as the first harmonic for a Fourier-synthesis of
lattice potentials, is produced with a multiphoton Raman
technique. The scheme uses three-level atoms with two stable
ground states $|g_0\rangle$ and $|g_1\rangle$ and one
electronically excited state $|e\rangle$. The atoms are irradiated
by two copropagating optical beams of frequency
$\omega+\Delta\omega$ and $\omega-\Delta\omega$ respectively and a
counterpropagating beam of frequency $\omega$, see
Fig.~\ref{fig2}b. During the induced virtual four-photon
processes, the optical field exchanges momentum with the atoms in
units of four photon recoils with the atoms, which is a factor two
above the corresponding processes in a standing wave. The spatial
periodicity of the induced four-photon lattice potential is
$\lambda/4$ \cite{Salger07}. In our experiment, the rubidium F = 1
ground state Zeeman sublevels $m_{\rm F} = -1$ and 0 are used as
levels $|g_0\rangle$ and $|g_1\rangle$, and the 5P$_{3/2}$
manifold as the excited state $|e\rangle$. A homogeneous magnetic
bias field of 1.8 G removes the degeneracy of the Zeeman
sublevels. The light to generate the optical lattice potential is
produced by a high power diode laser detuned 1 nm to the red of
the rubidium D$_2$ line. The emitted beam is split into two, from
which the two counter-propagating beams for generation of the
periodic potential is derived. In each of the beams, an
acoustooptic modulator generates all required optical frequencies
in the corresponding beam path. The potential depths and the
relative phase of the two spatial lattice harmonics can be
controlled individually.

\begin{figure}
\includegraphics{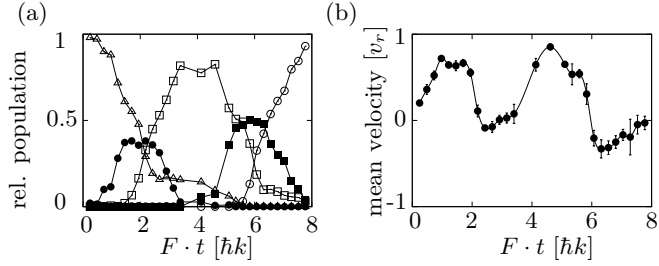}
\caption{%
Experimental data for Bloch-Zener oscillations of atoms in the
biharmonic lattice potential, which was accelerated with a
constant acceleration relatively to the atomic rest frame. (a)
Relative population of diffraction orders versus time: zeroth
order (triangles), first order (filled circles), second order
(open squares), third order (filled squares), and fourth order (open
circles) respectively. The solid line is to guide the eye. (b)
Mean atomic velocity (circles) in the co-moving frame versus time, 
where $v_r=\hbar k /m$ denotes the recoil velocity and $m$ the atomic mass.
The solid line is a spline fit.} \label{fig3}
\end{figure}

After preparation of the rubidium Bose-Einstein condensate, the
CO$_2$-laser trapping beam is extinguished, leaving the atoms in
ballistic free fall. The vertically oriented lattice beams are
activated 2.5 ms after the release of the condensate from the
trap, so that due to the lowering of the density during expansion 
the interatomic interactions are reduced. During the free fall of
the atoms, the lattice beams are initially switched on to
adiabatically load the atoms into the lowest band of the
Fourier-synthesized lattice at $q = 0$. For a measurement of
Bloch-Zener oscillations one of the lattice beams (the beam shown
on the right side of Figs.~\ref{fig2}a and ~\ref{fig2}b) is
subsequently acoustooptically detuned with a constant chirp-rate to
accelerate the lattice with respect to the atomic rest frame. Note
that a tuning of this beam's frequency moves both the $\lambda/2$
and the $\lambda/4$ spatial component in a phase-stable way, so
that the shape of the lattice is not affected by the acceleration
sequence. 
For the shown experimental data the depths of the lattice harmonics were 
$V_1 = 2.7$ $E_r$ and  $V_2 = 3.2$ $E_r$, where $E_r = (\hbar k)^2/2m$ denotes 
the recoil energy, the relative phase between the harmonics was 
$\phi = 0^\circ$.
The acceleration used in the atomic frame was 25 m/s$^2$. Fig.~\ref{fig3}a
shows the measured relative populations of the different
diffraction orders of the lattice after a variable acceleration
time $t$. The data allows to follow the wavepacket motion in the
miniband Bloch band structure (see Fig.~\ref{fig1}a) in time. When
$F\cdot t\approx\hbar k$, the band gap between the ground band and
the first excited Bloch band is reached, resulting a subsequent
decrease of the observed population of the zeroth diffraction
order (data with triangles) and an increase of the population of
the first diffraction order to roughly 40\% (filled circles),
which is attributed to the partial Landau-Zener tunneling into
the first excited Bloch band. Note that the splitting at this band
gap does not exactly match a 50/50 ratio. When $F\cdot
t\approx2\hbar k$, the zero's order peak is Bragg deflected into
the second order peak (data with open squares), i.e. only the
mapping onto the free eigenstates charges, while when $F\cdot
t\approx 3\hbar k$ is reached again a partial Landau-Zener
transition occurs with the relative phase between wavepackets
being such that most of the population is transferred into the
lowest energy band, which after crossing maps onto the second
diffraction order, and the observed corresponding relative
population then increases to roughly 85\%. 
We attribute the difference to 100\% as being mainly due to partial Landau-Zener 
tunneling into higher bands. 
For larger times, a second cycle of the wavepacket motion in the miniband 
structure is observed. We attribute the corresponding data as evidence for
Bloch-Zener oscillations in the Fourier-synthesized optical
lattice. Fig.~\ref{fig3}b gives the mean velocity of the atoms in
the co-accelerated frame versus time, as derived from the data of
Fig.~\ref{fig3}a.

\begin{figure}
\includegraphics{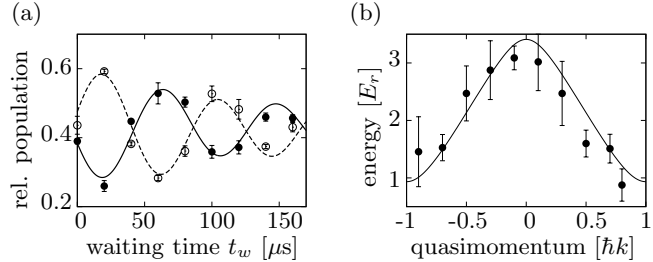}
\caption{%
(a) Measured St\u"ckelberg interference fringes based on two
partial Landau-Zener transitions between the bandgaps of the
optical lattice. The filled (open) circles give the measured relative
population in the first (second) diffraction respectively
versus the waiting time $t_w$, with $q_0=1.9$~$\hbar k$ for this
data set. The dashed and solid lines are fits to the data. (b) The
data points give the interferometrically determined energy
difference between the two minibands versus the lattice
quasimomentum in the atomic frame. The solid line is the
calculated energy difference.
}
\label{fig4}
\end{figure}

Fig.~\ref{fig4}a shows data where the relative phase was varied between the two
interfering paths of the formed Landau-Zener-St\"uckelberg band interferometer.
For the
corresponding measurement, the acceleration sequence was stopped after the first
beam splitting process at a certain value of the quasimomentum $q_0$
(with $q_0 = 1.9\hbar k$ for the here shown data) for a variable waiting time to
induce a variable phase shift. The filled (open) circles in Fig.~\ref{fig4}a give the
measured population in the first (second) order Bragg peak after releasing the
condensate from the lattice when reaching a quasimomentum of $3\hbar k$. As a
function of the waiting time $t_w$, which due to the energy difference of the
two bands tunes the accumulated relative phase, a clear interference pattern in
the relative population of Bloch bands is observed at the interferometer output.
From the measured oscillation frequency,
the energy difference between the Bloch bands at the corresponding quasimomentum
can readily be determined.
We attribute the observed damping of the fringe signal for large times $t_w$ 
to the finite velocity spread of atoms after their release from the trap. By 
the time of the experiment, the condensate interaction energy has been converted into 
kinetic energy. The corresponding velocity spread is expected to cause a nonzero 
width of the quasimomentum distribution in the lattice. This will lead to a 
spread of oscillation frequencies of the St\"uckelberg interference signal, and 
a corresponding loss of contrast for larger times $t_w$. 
The solid and dashed lines in Fig.~\ref{fig4}a are fits to the experimental 
fringe signal with a convolution of sinusoidal curves to account for the loss of 
contrast for larger times $t_w$ (for a measured atomic velocity spread 
of $\pm 0.8\hbar k$ for this data set).  
We note that in a quantum simulations view of our two-bands Bloch structure
experiment, the observed oscillation frequency can be interpreted as a beating
occurring at the Zitterbewegung frequency \cite{Vaishnav08, Gerritsma10}.
Fig.~\ref{fig4}b shows the derived energy splitting between the ground and the
first excited Bloch band of the miniband structure as a function of the
quasimomentum for the complete Brillouin zone. The solid line is the
theoretically expected curve of the energy difference for the given experimental 
parameters of the potential depths and relative phase of lattice harmonics, 
which is in good agreement with the corresponding experimental values. 
The smaller variation of the measured energy difference over the Brillouin zone 
with respect to the theoretical curve is attributed to the nonzero atomic 
velocity spread. 

To conclude, we have observed Bloch-Zener oscillations in an
optical lattice using a miniband-type Bloch band structure. The
relative phase of wavepackets in different Bloch bands was varied,
which has allowed to demonstrate the coherence of the
formed Landau-Zener interferometer and demonstrate a novel method
to interferometrically map out the energy difference between bands over the complete
Brillouin zone.

We expect that the described method has prospects for precise
interferometric determinations of the band structure in optical
lattices. 
The velocity spread of the atoms can be reduced by Raman selection 
\cite{Kasevich91}.
In principle, the method should also be applicable to
usual standing wave lattices when during the acceleration sequence
the Landau-Zener tunneling rate between Bloch bands is tailored dynamically 
either by appropriate variation of the potential depth of the lattice, similar 
as described in \cite{Zenesini09}, or by variation of the acceleration rate 
with time.
For example, when the ramp speed is correspondingly reduced in the vicinity 
of the crossing between the first two excited bands, the tunneling rate to 
higher bands can be kept small. Other prospects of the discussed coherent
manipulation in the miniband structure can include quantum
simulations of the (linear and nonlinear) Dirac equation
\cite{Gerritsma10}. The described interferometric method may also
be used to test for deviations of Newton's law over microscopic
distances, in a spirit similar than earlier experiments based on
Bloch-oscillations \cite{Ferrari06}.

We thank D. Witthaut for helpful discussions. Financial support from the
Deutsche Forschungsgemeinschaft is acknowledged.

%
\end{document}